\documentclass[letterpaper, 10 pt, conference]{ieeeconf}
\IEEEoverridecommandlockouts
\title{Semidefinite relaxations in optimal experiment design with application to substrate injection for hyperpolarized MRI}

\author{John Maidens and Murat Arcak
\thanks{J. Maidens and M. Arcak are with the Department of Electrical Engineering \& Computer Sciences,
University of California, Berkeley,
253 Cory Hall, Berkeley, CA, 94720 USA
e-mail: $\{$maidens, arcak$\}$@eecs.berkeley.edu. 
Research supported in part by NSERC postgraduate fellowship PGFD3-427610-2012. }}

\usepackage{comment}
\usepackage{amsfonts}
\usepackage{amsmath}
\usepackage{cite} 
\usepackage{graphicx}
\usepackage{algpseudocode}
\usepackage{algorithmicx}
\usepackage{algorithm}
\usepackage{subcaption}
\usepackage{placeins}
\usepackage[hyphens]{url}
\usepackage{hyperref} 
\usepackage{multirow}
\usepackage{flushend}

                               \renewcommand{\P}{\mathcal{P}}

                               \newcommand{\I}{\mathcal{I}}
                               
                               \newcommand{\R}{\mathbb{R}}

                               \newcommand{\tr}{\operatorname{trace}} 
                               \renewcommand{\u}{\mathbf{u}} 
                               \newcommand{\diag}{\operatorname{diag}}

\newtheorem{proposition}{Proposition}

\begin{document}
\thispagestyle{empty}
\pagestyle{empty}

\maketitle

\begin{abstract}
We consider the problem of optimal input design for estimating uncertain parameters in a discrete-time linear state space model, subject to simultaneous amplitude and $\ell_1$/$\ell_2$-norm constraints on the admissible inputs. We formulate this problem as the maximization of a (non-concave) quadratic function over the space of inputs, and use semidefinite relaxation techniques to efficiently find the global solution or to provide an upper bound. This investigation is motivated by a problem in medical imaging, specifically designing a substrate injection profile for \textit{in vivo} metabolic parameter mapping using magnetic resonance imaging (MRI) with hyperpolarized carbon-13 pyruvate. In the $\ell_2$-norm-constrained case, we show that the relaxation is tight, allowing us to efficiently compute a globally optimal injection profile. In the $\ell_1$-norm-constrained case the relaxation is no longer tight, but can be used to prove that the boxcar injection currently used in practice achieves at least 98.7\% of the global optimum. 
\end{abstract}

\section{Introduction} 

\subsection{Optimal experiment design} 

In this paper we consider the problem of estimating uncertain parameters in a state space model from noisy output data. For such a system, the reliability of the parameter estimates depends on the choice of input used to excite the system, as some inputs provide greater information about the parameters than others. The problem of designing an input that is maximally informative is known as optimal experiment design and much work has been done on this problem in the last 50 years \cite{Gevers11, Goodwin77, Ljung99, Walter97}.  

Historically, most work on optimal experiment design for dynamic systems has focused on frequency domain techniques, where an optimal input is designed based on its power spectrum. Here, we approach this problem in the time domain, allowing us to impose amplitude ($\ell_\infty$) and $\ell_1$ constraints on the admissible inputs. Amplitude-constrained optimal experiment design is NP-hard in general \cite{Manchester12}, but semidefinite relaxation techniques can be used to generate approximate solutions and to bound the suboptimality of such solutions \cite{Manchester12, Manchester10}. In contrast with \cite{Manchester12, Manchester10}, in this paper we restrict our attention to linear measures of the information allowing us to 1) write the objective function as a quadratic function of the input sequence and 2) apply an exactness result for quadratically constrained quadratic programming to give sufficient conditions under which the semidefinite relaxation recovers the global solution. We also present the results in terms of a state space model enabling us to model uncertainty in the initial state. 

\subsection{Metabolic MRI using hyperpolarized substrates} 

We are motivated by a problem in metabolic magnetic resonance imaging (MRI) using hyperpolarized substrates. 
Hyperpolarized carbon-13 MRI has enabled the real-time observation of perfusion and metabolism in preclinical and clinical studies \cite{Golman06, Day07, Kazan13, Nelson13, Bahrami14, Swisher14}. This technology is made possible by techniques for dynamic nuclear polarization (DNP) that have led to signal-to-noise ratio (SNR) increases of four to five orders of magnitude compared with endogenous signal in dissolved $^{13}$C-labelled molecules  \cite{Ardenkjaer-Larsen03, Golman03}.  Injected [1-$^{13}$C] pyruvate is frequently used as a substrate in metabolism experiments and its rate of conversion to  [1-$^{13}$C] lactate has been shown to distinguish between healthy and cancerous tissues in animal \cite{Day07}, and recently human \cite{Nelson13}, studies. 

The goal of a metabolic MRI experiment is to learn the spatial distribution of metabolic rates, as this indicates the regions of the body where a particular metabolic pathway is active. Noise in the observed image data leads to uncertainty in estimates of metabolic rate parameters, but the amount of uncertainty can be mitigated with experiment design. 

The problem of optimally designing an image acquisition sequence is considered in \cite{Maidens15ACC, Maidens15TMI}. In this paper we investigate optimal substrate injection profiles. After hyperpolarization the substrate must be injected into the test subject, where metabolism occurs and a sequence of images are acquired. Greater injection volumes lead to better signal-to-noise ratio, but for safety reasons the injection profile is limited by the rate at which substrate can be injected and by the total amount of fluid injected. Thus determining the optimal injection profile is of clinical interest. 

\subsection{Outline} 

We begin in Section \ref{sec:relaxation} by introducing the time domain optimal experiment design problem for linear systems, and discuss relevant relaxations from the literature. Then in Section \ref{sec:MRI} we present a mathematical model of hyperpolarized MRI and use semidefinite relaxation techniques to design optimal injection inputs for hyperpolarized substrates. We constrain the total amount of fluid injected using the $\ell_2$ or $\ell_1$ norms of the input signal. In the $\ell_2$-constrained case, the relaxation is tight and we are able to recover a globally optimal solution to the original problem. In the $\ell_1$-constrained case we do not recover a solution from the relaxation, but the objective value of the relaxation can be used to bound the optimality gap and show that a simple boxcar injection achieves at least 98.7\% of the global optimum. Matlab code to reproduce the results in this paper is available at \url{https://github.com/maidens/ACC-2016}.

\section{Semidefinite relaxations in optimal experiment design} 

\label{sec:relaxation} 
Consider a discrete-time linear system
\begin{equation}
\begin{split}
    x_{t+1} &= A(\theta) x_t + B(\theta) u_t \\
       y_t &= C x_t \\
    Y_t &\sim N(y_t, \Sigma) 
\end{split}
\label{eq:linear_system}
\end{equation}
over a horizon $0 \le t \le N$ where the initial state $x_0$ and time-invariant state dynamics matrices $A \in \R^{n \times n}$ and $B \in \R^{n \times n_u}$ are dependent on some unknown parameter vector $\theta \in \R^p$.  We wish to choose an input sequence $u$ to maximize some measure of the Fisher information matrix
\begin{equation}
   \I (\theta) = \sum_{t=0}^N (\nabla_\theta x_t)^T C^T \Sigma^{-1} C (\nabla_\theta x_t)
   \label{eq:Fisher} 
\end{equation} 
about the parameter vector $\theta$ contained in the data $Y$. The Fisher information is a positive semidefinite matrix that is used in experiment design for parameter estimation as a measure of the informativeness of an experiment \cite{Pukelsheim93, Walter97}. The inverse of the information gives an upper bound on the covariance of an arbitrary unbiased estimator $\theta$ via the Cramer-Rao inequality. See \cite{Maidens15Arxiv} for a derivation of (\ref{eq:Fisher}).

\subsection{Quadratic objective function}

We consider nonnegative, linear information metrics of the form 
\begin{equation} 
\varphi \big(\I(\theta)\big) = \tr \big( K \I(\theta) \big)
\label{eq:objective} 
\end{equation}
where $K \succeq 0$ is positive semidefinite. Particular cases include the T-optimal design criterion (where $K$ is the $p \times p$ identity matrix) \cite{Pukelsheim93} and the $c$-optimal design criterion (where $K = cc^T$) which is optimal for the scalar parameter $c^T \theta$ \cite{Pukelsheim81}. 

To formulate the problem of choosing the sequence $u_k$ for $k = 0, \dots, N-1$ we define the stacked vector $\u =  [u_0^T \dots u_{N-1}^T]^T$ and denote its components $\u^{(j, k)}$ where $(j, k) = k n_u + j$ defines a reverse lexicographic ordering, $j$ ranges over the input index from 1 to $n_u$ and $k$ ranges over the time index from 0 to $N-1$. With this notation established, we can state the following result. 
\begin{proposition} 
  The objective (\ref{eq:objective}) is a quadratic function 
  \[
  	\varphi\big(\I(\theta)\big) = \u^T Q(\theta) \u + 2 q(\theta)^T \u + q_0(\theta) 
  \]
  of the design variable $\u$. The entries of the matrices appearing in the objective function are computed as 
\begin{equation*}
  \resizebox{\columnwidth}{!}{$
  \begin{split}
  	Q(\theta)_{(j',k')(j, k)} &= \sum_{t = \max\{k, k'\} + 1}^N \sum_{h, h' = 1}^n \sum_{i, i' = 1}^p K^{ii'}M_{k'i'j'}^{h'}(t) S_{hh'}M_{kij}^h(t) \\
	q(\theta)_{(j, k)} &= \sum_{t = k + 1}^N \sum_{h, h' = 1}^n \sum_{i, i' = 1}^p K^{ii'} m_{i'}^{h'}(t) S_{hh'} M_{kij}^h(t) \\
	q_0(\theta) &=  \sum_{t=0}^N \sum_{h, h' = 1}^n \sum_{i, i' = 1}^p K^{ii'} m_{i'}^{h'}(t) S_{hh'} m_{i}^{h}(t)
  \end{split}
  	$}
  \end{equation*}
  where $S_{hh'}$ is the $h h'$-th entry of $C^T\Sigma^{-1}C$, $m_{i}^{h}(t)$ is the $h$-th entry of the vector \[ A^t \frac{\partial x_0}{\partial \theta_i} +  \sum_{\ell = 0}^{t-1} A^{t-\ell-1} \frac{\partial A}{\partial \theta_i} A ^\ell x_0\] and $M_{kij}^h(t)$ is the $h, j$-th entry of the matrix \[A^{t-k-1} \frac{\partial B}{\partial \theta_i} + \sum_{\ell = k+1}^{t-1} A^{t-\ell-1} \frac{\partial A}{\partial \theta_i} A^{\ell - k - 1} B.\] 
  \label{prop:quadratic}
\end{proposition} 

We see that for a linear dynamical system with Gaussian-distributed measurements, optimal experiment design is a nonconvex quadratic programming problem. If there are no constraints on the admissible inputs the optimal value is infinite, as by choosing $\u^*$ as an eigenvector of $Q(\theta)$ that corresponds to a nonzero eigenvalue, letting $\u = \alpha \u^*$ for $\alpha > 0$, we can make $ \u^T Q(\theta) \u + 2 q(\theta)^T \u + q_0(\theta) $ arbitrarily large. 

Constrained quadratic programming is NP-hard in general \cite{Garey79}. But certain quadratic programming problems lend themselves to polynomial-time approximation algorithms using a semidefinite programming relaxation \cite{Goemans95, Luo10}. We first give a proof of Proposition \ref{prop:quadratic}, before moving on to semidefinite relaxations of the optimal experiment design problem. 

\begin{proof} 
Unrolling the recursion relation (\ref{eq:linear_system}), we can write 
\[
  x_t = A ^t x_0 + \sum_{k = 0}^{t-1} A^{t-k-1} B u_k. 
\]
Applying the chain rule to (\ref{eq:linear_system}) we get a recursion relation for the sensitivities 
\[
  \frac{\partial}{\partial \theta_i} x_{t+1} = A \frac{\partial}{\partial \theta_i} x_t + \frac{\partial A}{\partial \theta_i} x_t +  \frac{ \partial B}{\partial \theta_i} u_t 
\]
which can be unrolled as 
\begin{equation*}
  \resizebox{\columnwidth}{!}{$
\begin{split}
   \frac{\partial}{\partial \theta_i} x_{t} 
   &= A^t \frac{\partial x_0}{\partial \theta_i} +  \sum_{\ell = 0}^{t-1} A^{t-\ell-1} \left[ \frac{\partial A}{\partial \theta_i} x_\ell +  \frac{ \partial B}{\partial \theta_i} u_\ell  \right]. \\
   &= A^t \frac{\partial x_0}{\partial \theta_i} +  \sum_{\ell = 0}^{t-1} A^{t-\ell-1} \left[ \frac{\partial A}{\partial \theta_i} \left( A ^\ell x_0 + \sum_{k = 0}^{\ell-1} A^{\ell-k-1} B u_k \right) +  \frac{ \partial B}{\partial \theta_i} u_\ell  \right] \\
   &= \left( A^t \frac{\partial x_0}{\partial \theta_i} +  \sum_{\ell = 0}^{t-1} A^{t-\ell-1} \frac{\partial A}{\partial \theta_i} A ^\ell x_0 \right) \\
  & \  + \sum_{k = 0}^{t-1} \left( A^{t-k-1} \frac{\partial B}{\partial \theta_i} + \sum_{\ell = k+1}^{t-1} A^{t-\ell-1} \frac{\partial A}{\partial \theta_i} A^{\ell - k - 1} B \right) u_k 
\end{split}
$}
\end{equation*}
To simplify notation, from this point forward we will use the Einstein summation convention, where upper and lower repeated indices denotes an implicit summation over those indices. In this notation, we see that the sensitivity matrix $\nabla_\theta x_t$ has entries 
\[
 (\nabla_\theta x_t)^h_i = M_{kij}^h(t) \u^{jk} + m_i^h(t) 
\]
where $M$ and $m$ are defined as in the statement of the proposition and the index $k$ runs from 0 to $t-1$. Thus the objective function is expressed as
\begin{equation*}
  \resizebox{\columnwidth}{!}{$
\begin{split} 
&\tr\Big(K \I(\theta)\Big) = K^{ii'} \I(\theta)_{ii'} \\
		&= K^{ii'} \sum_{t=0}^N  (\nabla_\theta x_t)^{h'}_{i'} S_{hh'}  (\nabla_\theta x_t)^h_i \\
		&= K^{ii'} \sum_{t=0}^N  \Big(M_{k'i'j'}^{h'}(t) \u^{j'k'} + m_{i'}^{h'}(t)\Big) S_{hh'}  \Big(M_{kij}^h(t) \u^{jk} + m_i^h(t)\Big) \\
		&= \sum_{k, k' = 0}^{N-1} \u^{j'k'} \left( \sum_{t = \max\{k, k'\} + 1}^{N} K^{ii'} M_{k'i'j'}^{h'}(t) S_{hh'} M_{kij}^h(t) \right) \u^{jk} \\
		& \ \ \ \ + 2 \sum_{k=0}^{N-1}\left( \sum_{t=k+1}^N K^{ii'} m_{i'}^{h'}(t) S_{hh'} M_{kij}^h(t) \right) \u^{jk}  \\
		& \ \ \ \ + \sum_{t=0}^N K^{ii'} m_{i'}^{h'}(t) S_{hh'} m_i^h(t) \\
		&= \u^T Q(\theta) \u + 2 q(\theta)^T \u + q_0(\theta) 
\end{split}
$}
\end{equation*} 
\end{proof}

\subsection{Semidefinite relaxation of quadratic programs } 


Semidefinite relaxations of indefinite quadratic programming problems  have been the subject of considerable study over the past two decades \cite{Goemans95, Luo10, Nesterov98, Nemirovski99, Ye99, Zhang00}. The essential idea is to take a quadratic program 
\begin{equation}
\begin{split}
   \textbf{maximize} & \ \ u^T Q u + 2q^T u + q_0\\ 
   \textbf{subject to} & \ \ u \in \P \subseteq \R^d \\
\end{split} 
\label{eq:QP} 
\end{equation} 
over a linear vector space and transform it to a linear problem over a quadratic space 
\begin{equation}
\begin{split}
   \textbf{maximize} & \ \ \tr\left( \left[ \begin{array}{cc}Q & q \\ q^T & q_0\end{array} \right] \left[ \begin{array}{cc}U & u\\ u^T & 1\end{array}\right]  \right) \\ 
   \textbf{subject to} & \ \ \left[ \begin{array}{cc}U & u\\ u^T & 1\end{array}\right]  \in \tilde\P \subseteq \R^{(d+1) \times (d+1)} \\ 
   		& \ \  U = uu^T 
\end{split} 
\label{eq:semidefinite_QP} 
\end{equation} 
by introducing a rank 1 semidefinite variable $U = uu^T$ and translating the constraint set $\P$ from the the linear vector space to a convex constraint set $\tilde \P$ in the semidefinite matrix space. The problem (\ref{eq:semidefinite_QP}) in non convex, but the equality constraint $U = uu^T$ can be relaxed to the inequality $U \succeq uu^T$ then transformed to semidefinite constraint $ \left[ \begin{array}{cc}U & u\\ u^T & 1\end{array}\right]  \succeq 0$ via Schur complement. Thus (\ref{eq:semidefinite_QP}) can be relaxed to the convex problem 
\begin{equation}
\begin{split}
   \textbf{maximize} & \ \ \tr\left( \left[ \begin{array}{cc}Q & q \\ q^T & q_0\end{array} \right] \left[ \begin{array}{cc}U & u\\ u^T & 1\end{array}\right]  \right) \\ 
   \textbf{subject to} & \ \ \left[ \begin{array}{cc}U & u\\ u^T & 1\end{array}\right]  \in \tilde\P \\ 
   		& \ \ \left[ \begin{array}{cc}U & u\\ u^T & 1\end{array}\right]  \succeq 0. 
\end{split} 
\label{eq:semidefinite_relaxation} 
\end{equation} 
The value of the convex program (\ref{eq:semidefinite_relaxation}) can then be used as an upper bound on the solution to (\ref{eq:QP}). Further if a solution $\left[ \begin{array}{cc}U^* & u^*\\ u^{*T} & 1\end{array}\right] $ to (\ref{eq:semidefinite_relaxation}) happens to have rank 1 then $u^*$ is a global solution to (\ref{eq:QP}).

To translate constraints $u \in \P \subseteq \R^d$ on $u$ to constraints $U \in \tilde \P \subseteq \R^{d \times d} $ on $U$ there are a number of methods. 
\begin{itemize}
\item Quadratic constraints on $u$ of the form \[\P = \{u: u^T R u + 2r^T u + r_0 \le 0\}\] are translated to constraints of the form \[\tilde \P = \{U : \tr \left(  \left[ \begin{array}{cc} R & r \\ r^T & r_0 \end{array} \right]  \left[ \begin{array}{cc} U & u \\ u^T & 1 \end{array} \right] \right) \le 0\}.\] For example, the $\ell_2$ constraint $\| u \|_2 \le c$ becomes $\tr(U) \le c^2$. 
\item Amplitude constraints on $u$ of the form $\P = \{u : | u_t | \le c_t \ t=1, \dots, d\}$ for particular constants $c_t$ are modelled as $d$ homogeneous quadratic constraints $\P = \{u : u_t^2 \le c_t^2 \ t=1, \dots, d\}$ which correspond to $\tilde \P = \{ U : U_{tt} \le c_t^2 \ t=1, \dots, d \}$. 
\item Box constraints of the form $\P = \{ u : 0 \le u_t \le c_t\}$ can be translated as $\tilde \P = \{ U : U_{tt} \le c_t^2 \ t=1, \dots, d\ \  \wedge \ \ U_{st} \ge 0 \ s, t = 1, \dots, d \}$. 
\item For $u$ normalized such that $0 \le u_t \le 1$, the relaxation can be tightened by noting that $u_i^2 \le u_i$. Thus the constraints $U_{ii} \le u_i \ \ i =1, \dots, d$ can be added to tighten the relaxation. 
\item If $u_t \ge 0$ and $a$ is a vector with $a_t \ge 0$ then the linear constraint $\P = \{u : a^T u \le b\}$ can be translated by noting that $0 \le a^T u \le b$ implies that $\tr(aa^T U) = \tr(a^T U a) = a^T u u^T a = (a^T u)^2 \le b^2$. This results in a tighter approximation than adding the constraint $a^T u \le b$ (see Lemma 1 of\cite{Helmberg96}). In particular, if we denote the $d \times d$ matrix of ones by $E$ then the $\ell_1$-norm constraint $\|u\|_1 \le b$ can be translated as $\tr(E U) \le b^2$. 
\end{itemize}  

\subsection{A result on exact recovery from semidefinite relaxation} 

For a vector $u\in \R^d$ let $u^2$ denote the vector obtained by squaring the entries of $u$ component-wise and consider a problem of the form 
\begin{equation}
\begin{split}
   \textbf{maximize} & \ \ u^T  Q u + 2q^T u + q_0 \\ 
   \textbf{subject to} & \ \ u^2 \in \mathcal{F}. \\
\end{split} 
\label{eq:exact_QP} 
\end{equation} 
The following result gives sufficient conditions for the semidefinite relaxation
\begin{equation}
\begin{split}
   \textbf{maximize} & \ \ \tr\left( \left[ \begin{array}{cc}Q & q \\ q^T & q_0\end{array} \right] \left[ \begin{array}{cc}U & u\\ u^T & 1\end{array}\right]  \right) \\ 
   \textbf{subject to} & \ \ \operatorname{diag}( U) \in \mathcal{F} \\
                                & \ \ \left[ \begin{array}{cc}U & u\\ u^T & 1\end{array}\right] \succeq 0
\end{split} 
\label{eq:exact_relaxation} 
\end{equation} 
to recover the global solution of (\ref{eq:exact_QP}). 

\begin{proposition}[Adapted from Theorem 2 of \cite{Zhang00}] 
	If $ Q_{ij} \ge 0$ for all $i \ne j$, $q_i \ge 0$ for all $i$  and $\mathcal{F} \subseteq \R^d$ is a closed convex set then the values of (\ref{eq:exact_QP})  and  (\ref{eq:exact_relaxation}) coincide. Moreover, if $\tilde U^*$ is a solution of (\ref{eq:exact_relaxation}) then $\sqrt{\diag(\tilde U^*)}$ is a solution of (\ref{eq:exact_QP}). 
	\label{prop:exact}
\end{proposition}

Thus if $u \in \P$ can be expressed in the form $u^2 \in \mathcal{F}$ for some convex $\mathcal{F}$, and the entries of $Q$ and $q$ are nonnegative, we can globally solve (\ref{eq:QP}) via the convex relaxation.

\section{Infusion input design for substrate injection in hyperpolarized carbon-13 MRI} 
\label{sec:MRI} 

We consider a linear model of magnetization exchange resulting from the injection of a hyperpolarized substrate, observed using a flip angle sequence $\alpha_{k, t}$ \cite{Khegai14}: 
\begin{equation}
  \resizebox{\columnwidth}{!}{$
   \frac{dx}{dt}(t) =  \left[ \begin{array}{cc}
   -k_{PL} - R_{1P} - \frac{1-\cos(\alpha_{1, t})}{\Delta t} & 0 \\ 
    k_{PL} & -R_{1L} -  \frac{1-\cos(\alpha_{2, t})}{\Delta t}   \end{array} \right] x(t) +  \left[ \begin{array}{c} k_{TRANS} \\ 0 \end{array} \right] AIF(t) 
    $}
    \label{eq:MRI}
\end{equation}
where $AIF(t)$ is an arterial input function. 
The result of a bolus (impulse) injection of substrate is often modelled as the arterial input function is of the form \cite{vonMorze11}
 \[
   AIF(t) = A_0 t^\gamma e^{-t/\beta} . 
 \]
 In the case $\gamma = 2$ the samples of this AIF can be modelled as the impulse response of the $\gamma+1 = 3$rd order system 
 \begin{equation}
   \resizebox{0.85\columnwidth}{!}{$
 \begin{split} 
 z_{t+1} &=  \left[ \begin{array}{ccc}
  3e^{-(\Delta t)/\beta} & -3 e^{-2(\Delta t)/\beta} & e^{-3(\Delta t)/\beta} \\
  1 & 0 & 0 \\
  0 & 1 & 0
   \end{array} \right] z_t +  \left[ \begin{array}{c} 1 \\ 0 \\ 0 \end{array} \right] u_t \\
   AIF_t &= A_0 \Big[ e^{-(\Delta t)/\beta} \ \ e^{-2(\Delta t)/\beta} \ \ 0 \Big] z_t.  
   \end{split} 
   $}
   \label{eq:AIF_state}
 \end{equation}  
We discretize (\ref{eq:MRI}) with step $\Delta t = 2$ s assuming a zero-order hold on the AIF, yielding a model 
 \begin{equation}
 \begin{split}
 \bar x_{t+1} &= \bar A\  \bar x_t+\bar B \ AIF_t \\
  y_t &= \left[ \begin{array}{ccccc}
0 & 0 & 0 & \sin(\alpha_{1, t}) & 0 \\
0 & 0 & 0 & 0 & \sin(\alpha_{2, t}) 
\end{array}\right] \bar x_t.
 \end{split} 
 \label{eq:ZOH}
 \end{equation}
 Combining (\ref{eq:AIF_state}) and (\ref{eq:ZOH}), we get a model of the full system (\ref{eq:infusion_ODE}) mapping the infusion input $u$ to the observed signals.  
 \begin{figure*}[!t]
 \normalsize 
 \begin{equation}
 \begin{split} 
   x_{t+1}&=  \left[ \begin{array}{ccccc}
   3e^{-(\Delta t)/\beta} & -3 e^{-2(\Delta t)/\beta} & e^{-3(\Delta t)/\beta} & 0 & 0\\
  1 & 0 & 0 & 0 & 0 \\
  0 & 1 & 0 & 0 & 0 \\
  A_0e^{-(\Delta t)/\beta}\bar B_1 & A_0e^{-2(\Delta t)/\beta}\bar B_1 & 0 & \bar A_{11} & \bar A_{12} \\ 
  A_0e^{-(\Delta t)/\beta}\bar B_2 & A_0e^{-2(\Delta t)/\beta}\bar B_2 & 0 & \bar A_{21} & \bar A_{22} \end{array} \right] x_t+  \left[ \begin{array}{c} 1 \\ 0 \\ 0 \\ 0 \\ 0 \end{array} \right] u_t \\
     y_t&= \left[ \begin{array}{ccccc}
0 & 0 & 0 & \sin(\alpha_{1, t}) & 0 \\
0 & 0 & 0 & 0 & \sin(\alpha_{2, t}) 
\end{array}\right] x_t
\end{split} 
\label{eq:infusion_ODE}
\end{equation}
\hrulefill
\vspace{4pt} 
\end{figure*} 


We now solve an example instance of this system with model parameters taken from \cite{Maidens15TMI} which are shown in Table \ref{tab:parameter_values} , along with noise covariance matrix $\Sigma = I$, horizon of $N = 30$ samples and a constant flip angle sequence $\alpha_{k, t} = 15^\circ$. Computation times for solving this problem are given in Table \ref{tab:computation}. 
\begin{table}[ht!]
\centering
   \resizebox{\columnwidth}{!}{
\begin{tabular}{ |cccccccc| }
\hline
   $R_{1P}$ & $R_{1L}$ &  $k_{PL}$ & $k_{TRANS}$ &  $t_0$ & $\gamma$ &  $\beta$ & $ A_0$                      \\
   1/10          &   1/10      &  0.07         & 0.055              & 3.2596  &  2.1430    &   3.4658 &  1.0411 $\times 10^4$ \\
   \hline
\end{tabular}
}
 \vspace{10pt}
\caption{Nominal parameter values used} 
\label{tab:parameter_values}
\end{table}

\begin{table*}[ht!]
\centering
\begin{tabular}{ |c|c|c|c|c| }
\hline
problem & time to generate SDP & SDP decision variable size & number of SDP constraints &  time to solve SDP \\
\hline
$\ell_2$ constrained  & \multirow{2}{*}{12.0 s} & 30 $\times$ 30 symmetric & 31 & 0.58 s \\
$\ell_1$ constrained & & 31 $\times$ 31 symmetric & 497 & 5.51 s \\
   \hline
\end{tabular}
 \vspace{10pt}
\caption{Computation time to solve semidefinite programming relaxations. The SDP was solved using CVX \cite{cvx} with the SeDuMi backend in MATLAB v8.4.0 running on a Macbook laptop (2.3 GHz quad-core Intel Core i7 Ivy Bridge processor, 8GB memory).} 
\label{tab:computation}
\end{table*}

\subsection{$\ell_2$ constrained input} 

\begin{figure}[ht!] 
    \centering
    \begin{subfigure}[b]{0.7\columnwidth}
        \includegraphics[width=\textwidth]{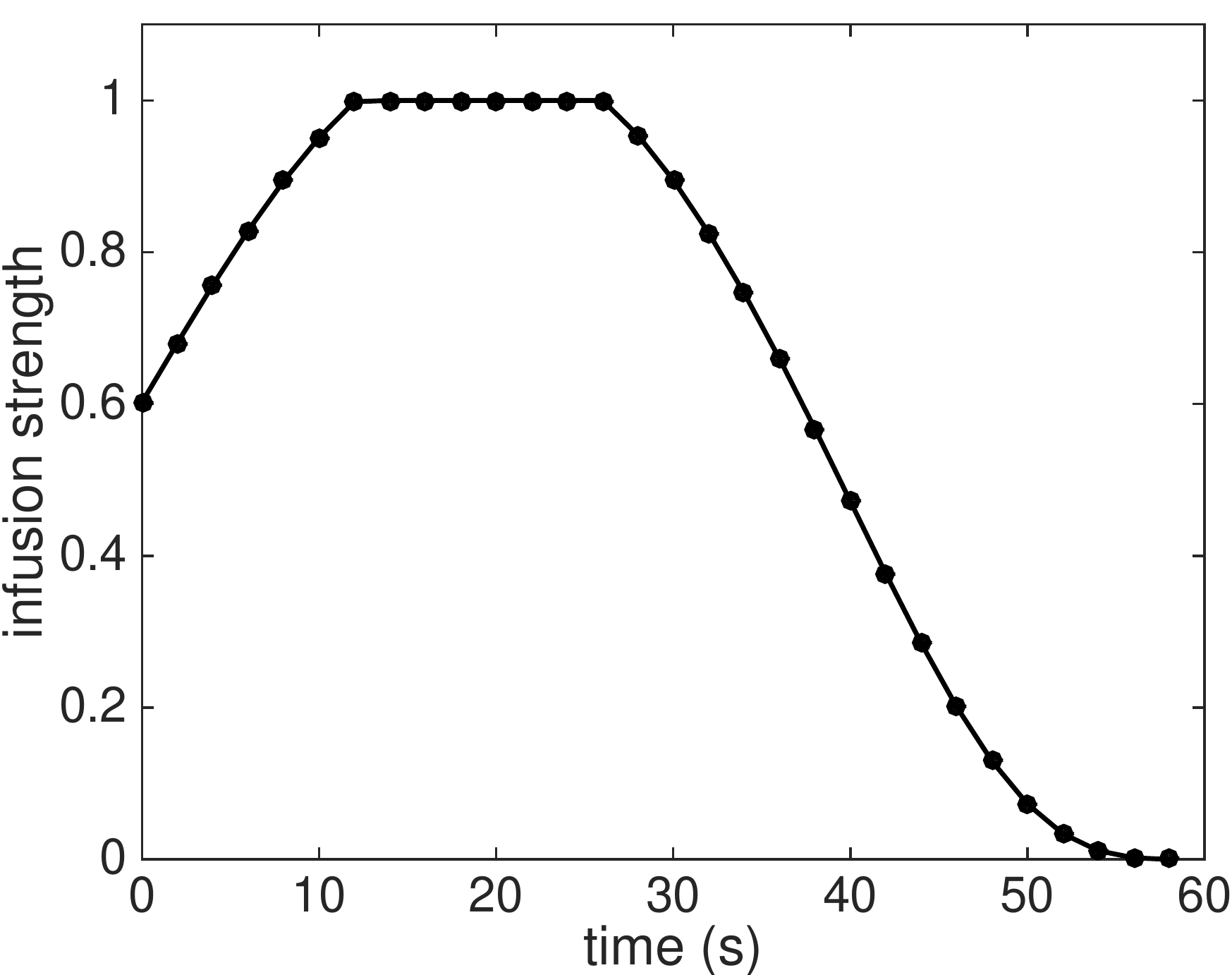}
        \caption{Optimal input}
    \end{subfigure}
    ~ 
    \begin{subfigure}[b]{0.7\columnwidth}
        \includegraphics[width=\textwidth]{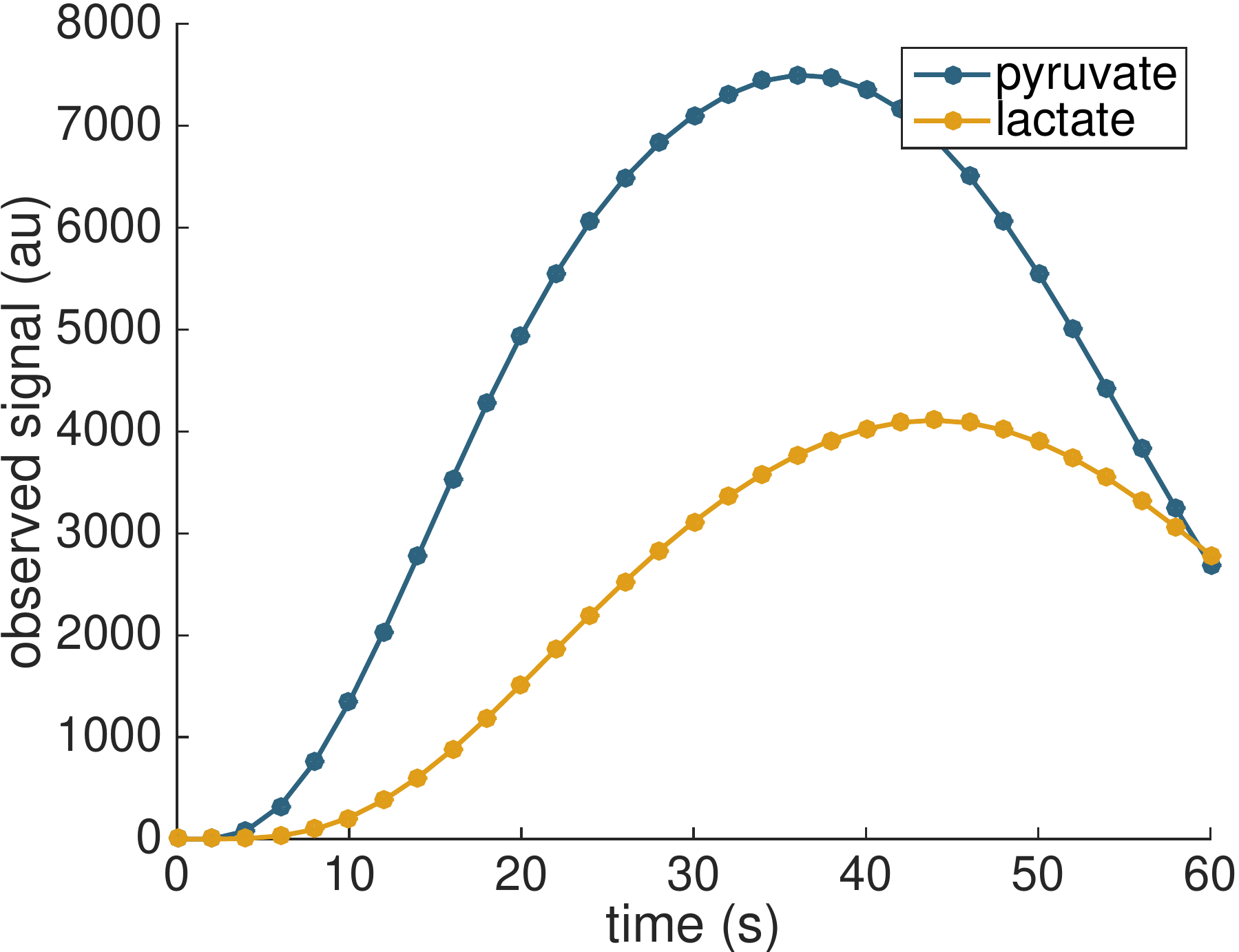}
        \caption{Corresponding outputs}
    \end{subfigure}
    \caption{Optimal solution to (\ref{eq:l2_program}) computed using semidefinite relaxation. }
 \label{fig:l2_solution} 
\end{figure}

The substrate injection is constrained to limit the rate of injection to $| u_t | \le 1$ and to limit the $\ell_2$ norm of the injection to $\| u \|_2 \le 4$. Thus we wish to solve the quadratically-constrained quadratic program
\begin{equation}
\begin{split}
   \textbf{maximize} & \ \  u^T Q(\theta) u \\ 
   \textbf{subject to} & \ \ u_t^2  \le 1 \\
                                & \ \ \sum_{t=0}^{N-1} u_t^2 \le 16.  
\end{split} 
\label{eq:l2_program} 
\end{equation}
The matrix $Q$ is nonnegative and the constraints are of the form $u^2 \in \mathcal{F}$ where $\mathcal{F}$ is a closed convex set. Therefore this problem yields a semidefinite relaxation 
\begin{equation}
\begin{split}
   \textbf{maximize} & \ \  \tr\Big( Q(\theta) U \Big)  \\ 
   \textbf{subject to} & \ \ U_{tt} \le 1 \\
                                & \ \ \tr(U) \le 16.  
\end{split} 
\label{eq:l2_program_relaxation} 
\end{equation}
whose solution has rank 1 (by Proposition \ref{prop:exact}). So we can extract a globally optimal solution to (\ref{eq:l2_program}) from its semidefinite relaxation. The resulting optimal input trajectory and the corresponding output trajectories are shown in Fig. \ref{fig:l2_solution}.

\subsection{$\ell_1$ constrained input} 

We now replace the $\ell_2$-norm constraint with an $\ell_1$-norm constraint $\|u\|_1 \le 8 $ that limits the total amount of substrate injected. This constraint is more clinically relevant than the $\ell_2$-norm constraint, as the total substrate than can be injected is often limited due to safety-related concerns. This leads to a linearly-constrained QP of the form 
\begin{equation}
\begin{split}
   \textbf{maximize} & \ \  u^T Q(\theta) u \\ 
   \textbf{subject to} & \ \ 0 \le u_t  \le 1 \\
                                & \ \ \sum_{t=0}^{N-1} u_t  \le 8.  
\end{split} 
\label{eq:l1_program} 
\end{equation}
This QP can be relaxed to the semidefinite program 
\begin{equation}
\begin{split}
   \textbf{maximize} & \ \  \tr \Big( Q(\theta) U \Big) \\ 
   \textbf{subject to} & \ \ \left[\begin{array}{cc} U & u \\ u^T & 1 \end{array} \right] \succeq 0  \\
   			       & \ \ U_{tt} \le u_t \\
                                & \ \ \tr \Big( E U \Big) \le 64 \\
                                & \ \ U_{st} \ge 0 
\end{split} 
\label{eq:l1_program_relaxation} 
\end{equation}
where both $U$ and $u$ are decision variables and $E$ is the $N \times N$ matrix of ones. In this case, the constraint set is more complex than in the $\ell_2$-norm constrained case. Therefore the solution to the semidefinite program does not have rank 1, so we cannot extract the solution to (\ref{eq:l1_program}). But the optimal value of (\ref{eq:l1_program_relaxation}) is an upper bound on the optimal value of (\ref{eq:l1_program}), so given a proposed solution of (\ref{eq:l1_program}) we can bound the optimality gap using the value of (\ref{eq:l1_program_relaxation}). 

\begin{figure}[ht!] 
    \centering
    \begin{subfigure}[b]{0.7\columnwidth}
        \includegraphics[width=\textwidth]{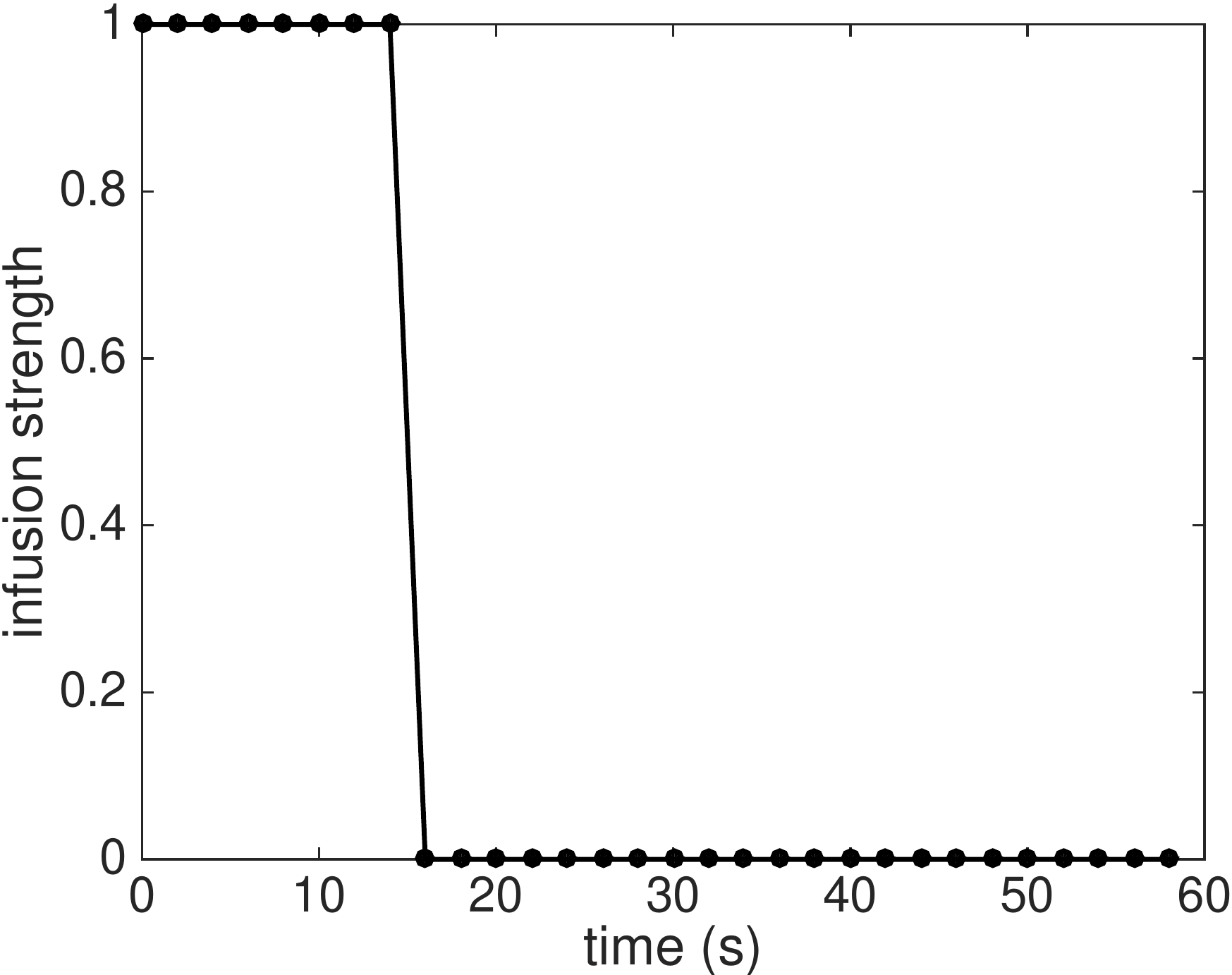}
        \caption{Boxcar input}
    \end{subfigure}
    ~ 
    \begin{subfigure}[b]{0.7\columnwidth}
        \includegraphics[width=\textwidth]{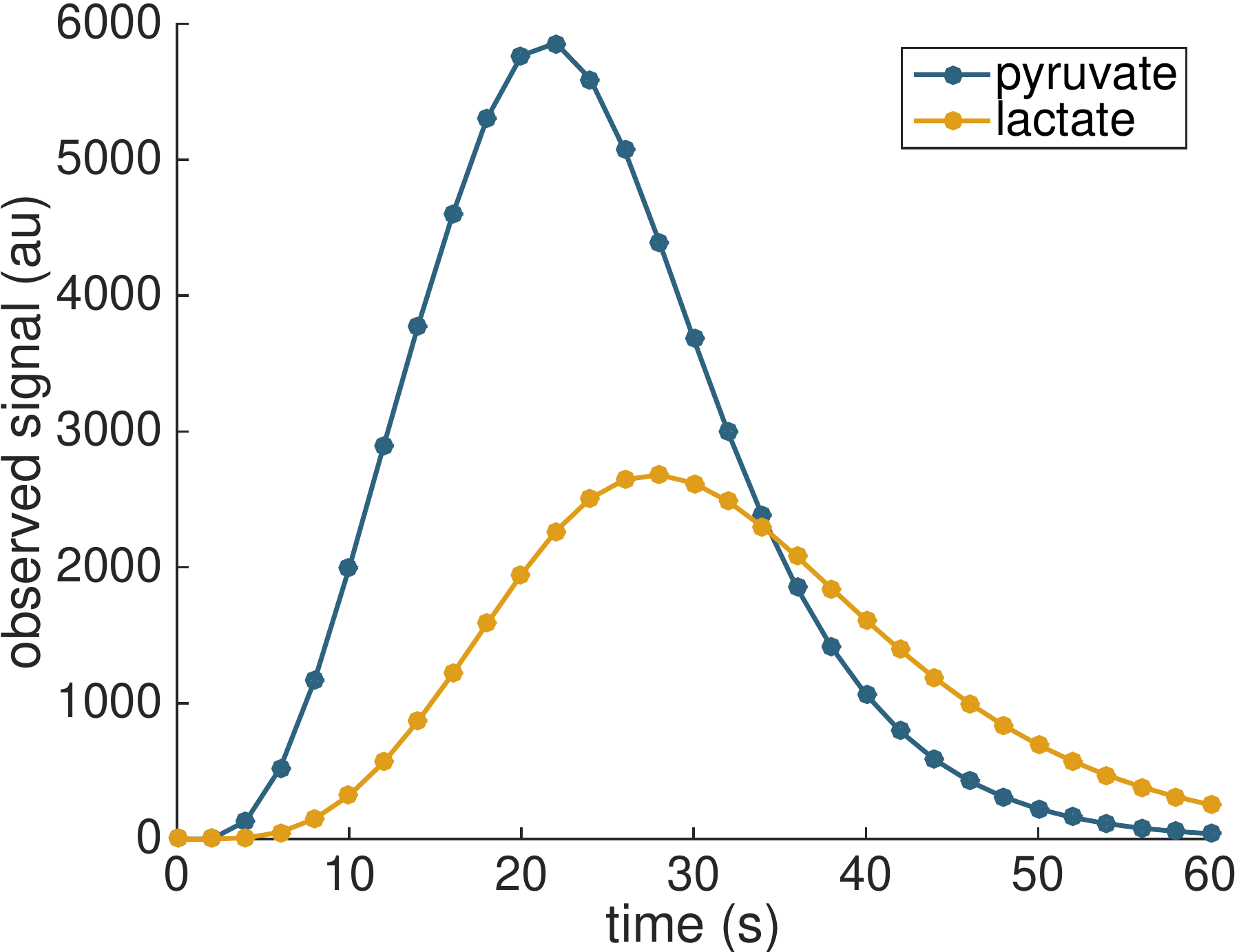}
        \caption{Corresponding outputs}
    \end{subfigure}
    \caption{The boxcar input function, which we conjecture to be optimal for the $\ell_1$-constrained problem. }
 \label{fig:boxcar} 
\end{figure}

We conjecture that the global solution of (\ref{eq:l1_program}) is the boxcar shown in Fig. \ref{fig:boxcar}. This input is what is currently used in practice: the substrate is injected at the maximum rate until the total allowable volume has been injected. Comparing the optimal value $1.2888 \times 10^{10}$ of the relaxation with the objective value $1.2724 \times 10^{10}$ we see that the boxcar input achieves a value of a factor of at least 0.9873 the optimal value. Thus even if our conjecture is incorrect and the boxcar is not optimal, the improvement that may be achieved by the optimal input is negligible. This observation helps to validate the current practice in hyperpolarized MRI. 

\section{Conclusions} 

We have found that semidefinite relaxation can be used to compute an optimal hyperpolarized substrate infusion input profile for estimating uncertain metabolic rate parameters in metabolic MRI. Future work will focus on investigating the relationship between the rank 1 recovery that we see in the $\ell_2$-norm-constrained case and properties of the dynamic system such as positivity and passivity, and attempting to extend these results to nonlinear measures of the information such as the D- E- and A-optimality criteria used in \cite{Manchester10}.

\bibliographystyle{IEEEtran}
\bibliography{references}

\end{document}